\documentclass[12pt]{article}
\usepackage{epsfig}
\usepackage{graphicx}
\usepackage{a4}
\usepackage{latexsym}
\usepackage{cite}

\usepackage{color}
\usepackage{colordvi}

\graphicspath{{Pics/}}
\DeclareGraphicsExtensions{.eps,.ps}

\usepackage{pslatex}
\usepackage[latin1]{inputenc}
\usepackage[T1]{fontenc}

\newcommand{\gsim}{\raisebox{-0.07cm}{$\:\:\stackrel{>}{{\scriptstyle
 \sim}}\:\: $} }

\def\vc{{V^{}_c}}
\def\nc{{N^{}_c}}

\def\nf{{n^{}_{\! f}}}

\def\lntwo{{\ln 2}}
\def\lnsqtwo{{\ln^2 2}}
\def\ztwo{{\zeta_2}}
\def\zthree{{\zeta_3}}
\def\tagXF4{{C_{F_4}}}

\def\shat{{\hat s}}
\def\mt{{m}}
\def\muf{{\mu^{}_f}}
\def\mur{{\mu^{}_r}}
\def\alphas{{\alpha_s}}

\def\fqqn{f_{q\bar{q}}^{(0)}}

\def\fggn{f_{gg}^{(0)}}

\def\fqqon{f_{q\bar{q}}^{(10)}}
\def\fqqoo{f_{q\bar{q}}^{(11)}}
\def\fqqtn{f_{q\bar{q}}^{(20)}}
\def\fqqto{f_{q\bar{q}}^{(21)}}
\def\fqqtt{f_{q\bar{q}}^{(22)}}

\def\fgqon{f_{gq}^{(10)}}
\def\fgqoo{f_{gq}^{(11)}}
\def\fgqtn{f_{gq}^{(20)}}
\def\fgqto{f_{gq}^{(21)}}
\def\fgqtt{f_{gq}^{(22)}}

\def\fggon{f_{gg}^{(10)}}
\def\fggoo{f_{gg}^{(11)}}
\def\fggtn{f_{gg}^{(20)}}
\def\fggto{f_{gg}^{(21)}}
\def\fggtt{f_{gg}^{(22)}}

\def\cqqtn{C_{x,q\bar{q}}^{(20)}}
\def\cgqtn{C_{x,gq}^{(20)}}
\def\cggtn{C_{x,gg}^{(20)}}

\def\cbqqtn{C_{\beta,q\bar{q}}^{(20)}}
\def\cbggtn{C_{\beta,gg}^{(20)}}

\def\Lij{{{\cal L}_{ij}}}
\def\fijn{{f_{ij}^{(0)}}}

\def\fijt{{f_{ij}^{(2)}}}

\def\fijon{f_{ij}^{(10)}}
\def\fijoo{f_{ij}^{(11)}}
\def\fijtn{f_{ij}^{(20)}}
\def\fijto{f_{ij}^{(21)}}
\def\fijtt{f_{ij}^{(22)}}

\begin{document}

\begin{titlepage}
\noindent
DESY 12-049 \hfill March 2012\\
HU-EP-12/10 \\
LTH 940\\
LPN 12-042 \\
SFB/CPP-12-16 \\
\vspace{1.3cm}

\begin{center}
  {\bf 
\Large
On top-pair hadro-production at next-to-next-to-leading order
  }
  \vspace{1.5cm}

  {\large
  S. Moch$^{\,a}$, 
  P. Uwer$^{\,b}$
  and A. Vogt$^{\, c}$}\\
  \vspace{1.2cm}

  {\it 
    $^a$Deutsches Elektronensynchrotron DESY \\
    Platanenallee 6, D--15738 Zeuthen, Germany
    \vspace{0.2cm}

    $^b$Humboldt-Universit\"at zu Berlin, Institut f\"ur Physik\\
    Newtonstra{\ss}e 15, D-12489 Berlin, Germany
    \vspace{0.2cm}

    $^c$Department of Mathematical Sciences, University of Liverpool \\
    Liverpool L69 3BX, United Kingdom
  }
  \vspace{2.4cm}

\large
{\bf Abstract}
\vspace{-0.2cm}
\end{center}
We study the QCD corrections at next-to-next-to-leading order (NNLO) 
to the cross section for the hadronic pair-production of top quarks. 
We present new results in the high-energy limit using 
the well-known framework of $k_t$-factorization. 
We combine these findings with the known threshold corrections 
and present improved approximate NNLO results over the full kinematic range. 
This approach is employed to quantify the residual theoretical uncertainty 
of the approximate NNLO results which amounts to about 
4\% for the Tevatron and 5\% for the LHC cross-section predictions.
Our analytic results in the high-energy limit will provide an important check on future
computations of the complete NNLO cross sections.
\vfill
\end{titlepage}

%
%
\newpage

The cross section for top-quark pair production has been measured
very precisely at the hadron colliders Tevatron and LHC 
with an experimental accuracy challenging the precision provided 
by the perturbative QCD corrections at next-to-leading order (NLO), 
which have been known for a long time~\cite{Nason:1987xz,Beenakker:1988bq}, 
see also~\cite{Bernreuther:2004jv,Czakon:2008ii}. 
Much recent activity has been concerned with improvements of the theoretical status beyond NLO, 
see~\cite{Bonciani:2012zt} and refs.~therein.
The dominant terms at higher orders have been used 
to derive approximate QCD corrections to next-to-next-to-leading order (NNLO) 
for the inclusive cross section~\cite{Moch:2008qy}.
These consist of large threshold logarithms at next-to-next-to-leading logarithmic accuracy (NNLL) 
which can even be resummed to all orders in perturbation theory 
and could provide sufficiently precise phenomenological predictions.
Yet, recent phenomenological studies based on threshold resummation 
to NNLL~\cite{Langenfeld:2009wd,Ahrens:2010zv,Kidonakis:2010dk,Beneke:2011mq,Cacciari:2011hy}
have reported somewhat differing predictions and, moreover, have proposed different means 
of estimating the residual theoretical uncertainty which is predominantly
due to uncalculated higher orders (beyond NNLO) and the effects of hard radiation 
not accounted for by threshold enhanced logarithms.

In this letter we consider the constraints on hadronic heavy-flavor production 
imposed by the high-energy factorization of the cross section~\cite{Catani:1990eg,Ball:2001pq}. 
This provides important complementary information on the hard partonic scattering processes 
in the limit when the center-of-mass energy is much larger than the top-quark mass.
It allows to extend previous approximations of the exact (yet unknown) NNLO results 
to the entire kinematical range and thus to obtain a more realistic uncertainty inherent 
in those approximate NNLO results.

The hadronic cross section for top-quark pair production is computed by 
the convolution of the partonic scaling functions $f_{ij}$ 
with the parton luminosities $\Lij$,
\begin{equation}
  \label{eq:totalcrs}
  \sigma_{h_1 h_2 \to {t\bar t X}}(S,\mt) = 
        {\alphas^2\over \mt^2}\,
        \sum_{i,j}
        \int\limits_{4\mt^2}^{S}\,
        ds \,\, \Lij(s, S, \mu)\,\,
        f_{ij}(s, \mt, \mu,\alphas) 
        \, , 
\end{equation}
where $S$ denotes the hadronic center-of-mass energy squared, 
and $\mt$ the top-quark mass in the on-shell (pole-mass) scheme. 
The parton luminosities $\Lij$ are defined as 
\begin{equation}
  \label{eq:partonlumi}
  \Lij(s,S,\mu) = 
  {1\over S} \int\limits_s^S
  {d\shat\over \shat} 
  f_{i/h_1}\left({\shat \over S},\mu\right) 
  f_{j/h_2}\left({s \over \shat},\mu\right)
  \, ,
\end{equation}
with the standard parton distribution functions (PDFs) $f_{i/h_{1,2}}(x,\mu)$.
The QCD coupling constant $\alphas$ is evaluated at the scale $\mu$ 
in the $\overline{MS}$ scheme with $\nf$ light flavors, and 
the renormalization and factorization have been identified (i.e., $\mu = \mur = \muf$).   
Up to NNLO, the scaling functions can be expanded as 
\begin{equation}
  \label{eq:DefinitionScalingFunctions}
  f_{ij} \,=\, 
  \fijn
  + 
  4\pi \alphas \left\{ \fijon +  L_M \fijoo \right\}
  +
  (4\pi \alphas)^2 \left\{ \fijtn + L_M \fijto + L_M^2 \fijtt \right\}
  \,+\, O(\alphas^3)
  \, ,
\end{equation}
where we abbreviate $L_M = \ln(\mu^2 / \mt^2)$. 
The dependence on $L_M$, included by the functions $\fijto$ and $\fijtt$ 
is known exactly from~\cite{Kidonakis:2001nj,Langenfeld:2009wd,Aliev:2010zk}.

For the high-energy factorization one considers Mellin moments $\omega$ 
with respect to $\rho = 4 \mt^2/s$, 
\begin{equation}
  \label{eq:mellindef}
  f_{ij}(\omega,\mu) \,=\,
  \int\limits_{0}^{1}\,d\rho\, \rho^{\omega-1}\,
  f_{ij}(\rho,\mu)
  \, .
\end{equation}
The resummation of the high-energy logarithms in $\rho$ for $\rho \to 0$,  
or, equivalently, of the singular terms in Mellin space as $\omega \to 0$, 
is based on the framework of PDFs un-integrated in the transverse momentum 
$k_t$ and the concept of $k_t$-factorization.
It is often also denoted to as small-$x$ resummation referring to the context of
deep-inelastic scattering (DIS).
The procedure involves two steps, i.e., the computation 
of amplitudes with the initial particles off-shell in $k_t$, 
and the subsequent convolution with a gluon PDF which has the 
corrections for small-$\rho$ included. 
For hadronic heavy-quark production, this leads to an expression  
for the cross section in Mellin space 
as a product of the (small-$x$ resummed) gluon PDF and the corresponding impact factor 
depending on $\omega$ through an anomalous dimension $\gamma_\omega$. 
which is determined by the well-known BFKL kernel.

For the purpose of this letter we are interested 
in the NNLO predictions of high-energy factorization in the framework of standard collinear factorization.
This requires the computation of the leading terms in Mellin space as $\omega \to 0$.
Using the heavy-quark impact factor of~\cite{Ball:2001pq}, 
the analytic result for inclusive heavy flavor hadro-production at NLO~\cite{Czakon:2008ii}, 
the {\sc FORM} routines of~\cite{Vermaseren:1998uu,Remiddi:1999ew}, 
and the {\sc PSLQ} algorithm as implemented in {\sc Maple} 
we arrive at the following expressions for the scaling functions at high energies for
a general SU$(\nc)$ gauge theory, where we define $\vc=\nc^2-1$.

At Born level we have up to order ${\cal O}(\omega^1)$, 
\begin{eqnarray}
\label{eq:fqq0}
\fqqn &=& 
       \pi \* \biggl(
         {1 \over 15}
       - {1 \over 15} \* {1 \over \nc^2}
       \biggr)
       + \omega \* \pi \* \biggl(
          - {77 \over 450}
          + {77 \over 450} \* {1 \over \nc^2}
          + \biggl\{
            {2 \over 15}
          - {2 \over 15} \* {1 \over \nc^2}
          \biggr\} \* \lntwo
          \biggr)
\, ,
\\
\label{eq:fgg0}
\fggn &=& 
       \pi \* \biggl(
         {4 \over 15} \* {\nc \over \vc}
       - {7 \over 18} \* {1 \over \nc \* \vc}
       \biggr)
       + \omega  \* \pi \* \biggl(
          - {781 \over 900} \* {\nc \over \vc}
          + {43 \over 36} \* {1 \over \nc \* \vc}
          + \biggl\{
            {8 \over 15} \* {\nc \over \vc}
          - {7 \over 9} \* {1 \over \nc \* \vc}
          \biggr\} \* \lntwo
          \biggr)
\, .
\qquad
\end{eqnarray}
Note that subleading terms in $\omega$, i.e.. ${\cal O}(\omega^0)$ and higher 
are not predicted by BFKL evolution. 
These terms are however required for the asymptotic behavior in NNLO.

At NLO up to order ${\cal O}(\omega^0)$ 
with $\nf$ denoting the number of light flavors the functions read,  
\begin{eqnarray}
\label{eq:fqq10}
4\pi \fqqon &=& 
            {191 \over 5400} \* \nc
          - {839 \over 8100} \* {1 \over \nc}
          + {221 \over 3240} \* {1 \over \nc^3}
          - \biggl\{
            {2 \over 15} \* {1 \over \nc}
          - {2 \over 15} \* {1 \over \nc^3}
          \biggr\} \*  \ztwo
          + \biggl\{
            {1 \over 50}
          - {1 \over 50} \* {1 \over \nc^2}
          \biggr\} \* \nf
\ ,
\qquad
\\
\label{eq:fqq11}
4\pi \fqqoo &=&
            {11 \over 90} \* \nc
          - {11 \over 90} \* {1 \over \nc}
          - \biggl\{
            {1 \over 45}
          - {1 \over 45} \* {1 \over \nc^2}
          \biggr\} \* \nf
\, , 
\\
\label{eq:fgq10}
4\pi \fgqon &=& 
         {1 \over \omega}  \*  \biggl(
            {77 \over 225}
          - {41 \over 108} \* {1 \over \nc^2}
          \biggl)
       - {194893 \over 108000}
       + {131357 \over 64800} \* {1 \over \nc^2}
       + \biggl\{ 
         {154 \over 225}
       - {41 \over 54} \* {1 \over \nc^2}
       \biggr\} \* \lntwo
\ ,
\\
\label{eq:fgq11}
4\pi \fgqoo &=& 
         {1 \over \omega}  \*   \biggl(
          - {2 \over 15}
          + {7 \over 36} \* {1 \over \nc^2}
          \biggr)
          + {941 \over 1800}
          - {527 \over 720} \* {1 \over \nc^2}
          - \biggl\{
            {4 \over 15}
          - {7 \over 18} \* {1 \over \nc^2}
          \biggr\} \*  \lntwo
\, , 
\\
\label{eq:fgg10}
4\pi \fggon &=& 
         {1 \over \omega}  \*  \biggl(
            {308 \over 225} \* {\nc^2 \over \vc}
          - {41 \over 27} \* {1 \over \vc}
          \biggr)
          + {364751 \over 15120} \* {1 \over \vc} 
          - {6971 \over 1680} \* {1 \over \nc^2 \* \vc}
          - {736427 \over 108000} \* {\nc^2 \over \vc} 
\nonumber\\ &&
          + \biggl\{
            {616 \over 225} \* {\nc^2 \over \vc} 
          - {82 \over 27} \* {1 \over \vc} 
            \biggr\} \* \lntwo
          + 
            {8 \over 15} \* {\nc^2 \over \vc}
            \* \ztwo
          - \biggl\{ 
            {11 \over 20} \* {\nc^2 \over \vc}
          + {489 \over 35} \* {1 \over \vc} 
          - {141 \over 35} \* {1 \over \nc^2 \* \vc} 
            \biggr\} \* \zthree
\nonumber\\ &&
          + {8 \over 9} \* {\nc^2 \over \vc} \* \tagXF4
       + 
            {1 \over 720} \* {\nc \over \vc} \* \nf
\, ,
\\
\label{eq:fgg11}
4\pi \fggoo &=& 
         {1 \over \omega}  \*  \biggl(
          - {8 \over 15} \* {\nc^2 \over \vc}
          + {7 \over 9} \* {1 \over \vc}
          \biggr)
          + {407 \over 150} \* {\nc^2 \over \vc}
          - {103 \over 27} \* {1 \over \vc}
          - \biggl\{
            {16 \over 15} \* {\nc^2 \over \vc}
          - {14 \over 9} \* {1 \over \vc}
          \biggr\} \*  \lntwo
\, .
\end{eqnarray}

Finally, at NNLO we have up to order ${\cal O}(\omega^{-1})$,
\begin{eqnarray}
\label{eq:f20qq}
(4\pi)^2 \fqqtn &=& 
         {1 \over \omega^2}  \* {1 \over \pi} \* \biggl(
            {2462 \over 3375} \* \nc
          - {88463 \over 81000} \* {1 \over \nc}
          + {235 \over 648} \* {1 \over \nc^3}
          - \biggl\{
            {1 \over 15} \* \nc
          + {11 \over 360} \* {1 \over \nc}
          - {7 \over 72} \* {1 \over \nc^3}
          \biggr\} \* \ztwo
          \biggr)
\nonumber\\ &&
         + {1 \over \omega}  \*  \cqqtn
\ ,
\\
\label{eq:f21qq}
(4\pi)^2 \fqqto &=& 
         {1 \over \omega^2}  \* {1 \over \pi} \* \biggl(
          - {77 \over 225} \* \nc
          + {1949 \over 2700} \*  {1 \over \nc}
          - {41 \over 108} \*  {1 \over \nc^3}
          \biggr)
         + {1 \over \omega}  \* {1 \over \pi} \*  \biggl(
            {222613 \over 108000} \* \nc
          - {708437 \over 162000} \*  {1 \over \nc}
\nonumber\\ &&
          + {149807 \over 64800} \*  {1 \over \nc^3}
          - \biggl\{
            {154 \over 225} \* \nc
          - {1949 \over 1350} \*  {1 \over \nc}
          + {41 \over 54} \*  {1 \over \nc^3}
          \biggr\} \*  \lntwo
          - \biggl\{
            {1 \over 27} \* \nc
          - {2 \over 27} \*  {1 \over \nc}
\nonumber\\ &&
          + {1 \over 27} \*  {1 \over \nc^3}
          \biggr\} \*  \nf
         \biggr)
\, ,
\\
\label{eq:f22qq}
(4\pi)^2 \fqqtt &=& 
         {1 \over \omega^2}  \* {1 \over \pi} \*  \biggl(
            {1 \over 15} \* \nc
          - {59 \over 360} \*  {1 \over \nc}
          + {7 \over 72} \*  {1 \over \nc^3}
          \biggr)
         - {1 \over \omega}  \* {1 \over \pi} \*  \biggl(
            {1121 \over 3600} \* \nc
          - {2701 \over 3600} \*  {1 \over \nc}
          + {79 \over 180} \*  {1 \over \nc^3}
\nonumber\\ &&
          - \biggl\{
            {2 \over 15}  \* \nc
          - {59 \over 180}  \*  {1 \over \nc}
          + {7 \over 36}  \*  {1 \over \nc^3}
          \biggr\} \*  \lntwo
         \biggr)
\, ,
\\
\label{eq:f20gq}
(4\pi)^2 \fgqtn &=& 
         {1 \over \omega^2}  \* {1 \over \pi} \*  \biggl(
            {2462 \over 1125} \* \nc
          - {479 \over 324} \*  {1 \over \nc}
          - \biggl\{
            {2 \over 15} \* \nc
          + {7 \over 36} \*  {1 \over \nc}
          \biggr\} \* \ztwo
          \biggr)
         + {1 \over \omega}  \*  \cgqtn
\, ,
\\
\label{eq:f21gq}
(4\pi)^2 \fgqto &=&
         {1 \over \omega^2}  \* {1 \over \pi} \*  \biggl(
          - {77 \over 75} \* \nc
          + {41 \over 36} \*  {1 \over \nc}
          \biggr)
         + {1 \over \omega} \* {1 \over \pi} \*  \biggl(
            {1496933 \over 216000} \* \nc
          - {3625007 \over 226800} \* {1 \over \nc}
          + {6971 \over 3360} \* {1 \over \nc^3}
\nonumber\\ &&
          - \biggl\{
            {154 \over 75} \* \nc
          - {41 \over 18} \* {1 \over \nc}
          \biggr\} \* \lntwo
          - 
            {4 \over 15} \* \nc
          \* \ztwo
          + \biggl\{
            {11 \over 40} \* \nc
          + {489 \over 70} \* {1 \over \nc}
          - {141 \over 70} \* {1 \over \nc^3}
          \biggr\} \* \zthree
\nonumber\\ &&
          - {4 \over 9} \* \nc \* \tagXF4
          - \biggl\{
            {293 \over 7200}
          - {1 \over 54} \* {1 \over \nc^2}
          \biggr\} \* \nf
         \biggr)
\, ,
\\
\label{eq:f22gq}
(4\pi)^2 \fgqtt &=&
         {1 \over \omega^2}  \* {1 \over \pi} \*  \biggl(
            {1 \over 5} \* \nc
          - {7 \over 24} \*  {1 \over \nc}
          \biggr)
         + {1 \over \omega} \*  {1 \over \pi} \*   \biggl(
          - {1541 \over 1200} \* \nc
          + {7871 \over 4320} \*  {1 \over \nc}
          + \biggl\{
            {2 \over 5} \* \nc
          - {7 \over 12} \*  {1 \over \nc}
          \biggr\} \*  \lntwo
\nonumber\\ &&
          + \biggl\{
            {1 \over 45}
          - {7 \over 216} \*  {1 \over \nc^2}
          \biggr\} \*  \nf
         \biggr)
\, ,
\\
\label{eq:f20gg}
(4\pi)^2 \fggtn &=& 
         {1 \over \omega^2}  \* {1 \over \pi} \*  \biggl(
            {3089 \over 2250} \* {\nc \over \vc}
          + {19696 \over 3375} \* \nc
          - \biggl\{
            {59 \over 90} \* {\nc \over \vc}
          + {4 \over 15} \* \nc
          \biggr\} \*  \ztwo
          \biggr)
         + {1 \over \omega}  \*  \cggtn
\, ,
\\
\label{eq:f21gg}
(4\pi)^2 \fggto &=& 
         {1 \over \omega^2}  \* {1 \over \pi} \*  \biggl(
          - {616 \over 225} \* {\nc^3 \over \vc}
          + {82 \over 27} \* {\nc \over \vc}
          \biggr)
         + {1 \over \omega}  \* {1 \over \pi} \*  \biggl(
            {358409 \over 18000} \* {\nc^3 \over \vc}
          - {1252103 \over 22680} \* {\nc \over \vc}
          + {6971 \over 840} \* {1 \over \nc \* \vc}
\nonumber\\ &&
          - \biggl\{
            {1232 \over 225} \* {\nc^3 \over \vc}
          - {164 \over 27} \* {\nc \over \vc}
            \biggr\} \* \lntwo
          - 
            {16 \over 15} \* {\nc^3 \over \vc}
            \* \ztwo
          + \biggl\{
            {11 \over 10} \* {\nc^3 \over \vc}
          + {978 \over 35} \* {\nc \over \vc}
          - {282 \over 35} \* {1 \over \nc \* \vc}
            \biggr\} \* \zthree
\nonumber\\ &&
          - {16 \over 9} \* {\nc^3 \over \vc} \* \tagXF4
          - \biggl\{
            {293 \over 1800} \* {\nc^2 \over \vc}
          - {26 \over 75} \* {1 \over \vc}
          + {103 \over 324} \* {1 \over \nc^2 \* \vc}
            \biggr\} \* \nf
         \biggr)
\, ,
\\
\label{eq:f22gg}
(4\pi)^2 \fggtt &=& 
         {1 \over \omega^2}  \* {1 \over \pi} \*  \biggl(
            {8 \over 15} \* {\nc^3 \over \vc}
          - {7 \over 9} \* {\nc \over \vc}
          \biggr)
         + {1 \over \omega}  \* {1 \over \pi} \*  \biggl(
          - {1771 \over 450} \* {\nc^3 \over \vc}
          + {403 \over 72} \* {\nc \over \vc}
\nonumber\\ &&
          + \biggl\{
            {16 \over 15} \* {\nc^3 \over \vc}
          - {14 \over 9} \* {\nc \over \vc}
          \biggr\} \*  \lntwo
          + \biggl\{
            {4 \over 45} \* {\nc^2 \over \vc}
          - {47 \over 270} \* {1 \over \vc}
          + {7 \over 108} \* {1 \over \nc^2 \* \vc}
          \biggr\} \*  \nf
         \biggr)
\, ,
\end{eqnarray}
where $\zeta_i$ denote the values of the Riemann zeta-function and the
constant $\tagXF4$ is calculated from 
\begin{equation}
\label{eq:cxF4}
\tagXF4 
\,=\, \int\limits_{0}^{1}\,{d\rho \over \rho}\, F_4(x) 
\,=\, -0.1333
\, ,
\end{equation}
where $F_4(x)$ is given in eq.~(19) of~\cite{Czakon:2008ii} 
and the value for $\tagXF4$ has been determined numerically.

All of the above formulae may be easily converted to momentum space with 
the replacements $1/\omega^2 \to -\ln \rho$ and $1/\omega \to {\rm const}_\rho$, cf. eq.~(\ref{eq:mellindef}).
At NNLO, the leading terms (LL$_x$) proportional to $1/\omega^2$ in the NNLO
quantities $\fijt$ follow directly from~\cite{Ball:2001pq}.
In addition, the new next-to-leading terms (NLL$_x$) proportional to $1/\omega$
in the scale dependent parts $\fijto$ in $\fijtt$ have been derived using standard
renormalization group methods, see~\cite{Kidonakis:2001nj,Langenfeld:2009wd,Aliev:2010zk}. 
This leaves the unknown NLL$_x$ terms denoted by $\cqqtn$, $\cgqtn$ and $\cggtn$ in eqs.~(\ref{eq:f20qq}), (\ref{eq:f20gq}) and (\ref{eq:f20gg}).
It is a general feature of small-$x$ expansions that the formally subleading
terms are numerically important, and that the ratio of NLL$_x$ to the LL$_x$ term is large, 
see, e.g., eqs.~(\ref{eq:f21qq}), (\ref{eq:f21gq}) and (\ref{eq:f21gg}).
Therefore, an estimate for these unknown terms is phenomenologically required.

We estimate $\cqqtn$, $\cgqtn$ and $\cggtn$ as follows. 
It has been observed (and also exploited constructively)~\cite{Catani:1990eg} 
that the impact factors in the high energy factorization 
for a number of different processes with initial state hadrons are related to each other.
In particular, the Abelian part of the impact factor for heavy-quark hadro-production 
is connected by a simple rescaling proportional to $\nc$ 
from the one for heavy-quark DIS evaluated at the scale of $Q^2 = \mt^2$ for the photon virtuality.

In the latter case, that is for the deep-inelastic production of a heavy-quark pair via 
scattering off a virtual photon off an initial quark or gluon, 
the NLL$_x$ terms at NNLO have recently been addressed in~\cite{Kawamura:2012cr}. 
In DIS the heavy-quark coefficient functions are subject to an exact factorization~\cite{Buza:1995ie} 
in the asymptotic limit $Q^2 \gg \mt^2$ 
into the respective coefficient functions with massless quarks and heavy-quark
operator matrix elements (OMEs). 
The approximate NNLO results for those heavy-quark coefficient functions 
are based on the three-loop results of~\cite{Vermaseren:2005qc,Bierenbaum:2009mv} 
and can be extended to good accuracy to all scales for the photon virtuality, 
in particular also to the scale $Q^2 = \mt^2$, see~\cite{Kawamura:2012cr} for details.
We can use this information to estimate the ratios $r_{x,gg}$ and $r_{x,gq}$ of the NLL$_x$ to the LL$_x$ 
terms for $\fggtn$ and $\fgqtn$ in eqs.~(\ref{eq:f20gq}), (\ref{eq:f20gg}). 
Subsequently, we multiply these ratios with the exact LL$_x$ terms of
eqs.~(\ref{eq:f20gq}), (\ref{eq:f20gg}) which assumes, of course, that the non-Abelian contributions 
to the NLL$_x$ terms for heavy-quark hadro-production do not lead to
significant deviations.
This assumption is motivated by the fact that the LL$_x$ terms of the scaling functions at high energy 
are related by simple replacements of color factors, 
e.g., $\fggon = 4 \nc^2/\vc \fgqon$ to LL$_x$ accuracy. 
Also, in cases where the NLL$_x$ are known exactly, e.g., the three-loop splitting functions~\cite{Vogt:2004mw},
such relations still hold to a good approximation.
In this way we arrive at,
\begin{eqnarray}
\label{eq:cgq20}
\cgqtn &=& r_{x,gq}\,
       {1 \over \pi} \*  \biggl(
            {737813 \over 121500}
          - {251 \over 540}\*\ztwo
       \biggr)
\quad\quad \mbox{with} \quad r_{x,gq} = -5.6,\, \dots,\, -7.7
\, ,
\\
\label{eq:cgg20}
\cggtn &=& r_{x,gg}\,
       {1 \over \pi} \*  \biggl(
         {324403 \over 18000}
       - {251 \over 240}\*\ztwo
       \biggr)
\quad\quad \mbox{with} \quad r_{x,gg} = -4.8,\, \dots,\, -8.2
\, ,
\end{eqnarray}
where the terms in brackets derive from the LL$_x$ term of eqs.~(\ref{eq:f20gq}), (\ref{eq:f20gg}) 
proportional to $1/\omega^2$ with $\nc = 3$ and $\vc = 8$ substituted. 
The uncertainty ranges in the estimates for $r_{x,gq}$ and $r_{x,gg}$ from~\cite{Kawamura:2012cr} are mainly driven 
by the finite number of Mellin moments currently available for the heavy-quark OMEs~\cite{Bierenbaum:2009mv}, 
which limit the extrapolation to $Q^2=\mt^2$.
For $\cggtn$ in eq.~(\ref{eq:cgg20}) these findings are also corroborated by the results of a Pad{\'e} estimate.
See e.g., \cite{Ellis:1997sb} for definitions and the use of Pad{\'e} estimates at higher orders in perturbations theory.
We use eqs.~(\ref{eq:fgg0}), (\ref{eq:fgg10}) as input for a $[0,1]$ Pad{\'e}
estimate of $\fggtn$ to derive the value of $r_{x,gg} = -5.1$ and 
we have also checked that the Pad{\'e} procedure predicts the NLL$_x$ terms of
$\fgqto$, $\fgqtt$, $\fggto$ and $\fggtt$ 
in eqs.~(\ref{eq:f21gq}), (\ref{eq:f22gq}), (\ref{eq:f21gg}) and (\ref{eq:f22gg}) even with an accuracy of 5\%.

For $\fqqtn$ we can neither establish directly a relation to heavy-quark DIS nor can we 
perform a Pad{\'e} estimate due to the vanishing NLO limit.
Therefore, we use the same range of values for the ratio $r_{x,gg}$ given in eq.~(\ref{eq:cgg20}), 
however rescaled a factor $1.6$ derived from the respective ratios of the NLL$_x$ to the LL$_x$ terms 
for $\fggn$ and $\fqqn$ in eqs.~(\ref{eq:fqq0}), (\ref{eq:fgg0}).
The motivation for this procedure is again, the above mentioned relations 
of the various scaling functions under simple exchange of color factors, 
see~\cite{Catani:1990eg,Ball:2001pq}.
Thus we use
\begin{eqnarray}
\label{eq:cqq20}
\cqqtn &=& r_{x,qq}\,
       {1 \over \pi} \*  \biggl(
            {502417 \over 273375}
          - {251 \over 1215}\*\ztwo
       \biggr)
\quad\quad \mbox{with} \quad r_{x,qq} = -3.0,\, \dots,\, -5.1
\, ,
\end{eqnarray}
where the brackets contain the LL$_x$ result of eq.~(\ref{eq:f20qq}) 
with the substitution $\nc = 3$ and $\vc = 8$.
As a check, we note that this procedure, if applied to the above mentioned 
Pad{\'e} estimate for $\fggto$ and $\fggtt$ predicts the NLL$_x$ terms in $\fqqto$ and $\fqqtt$ 
of eqs.~(\ref{eq:f21qq}), (\ref{eq:f22qq}) again with an accuracy of typically 5\%.
Therefore, we conclude that the range for $r_{x,qq}$ quoted in 
eq.~(\ref{eq:cqq20}) is a rather conservative one.

\begin{figure}[th!]
\centering
    {
    \includegraphics[width=7.25cm]{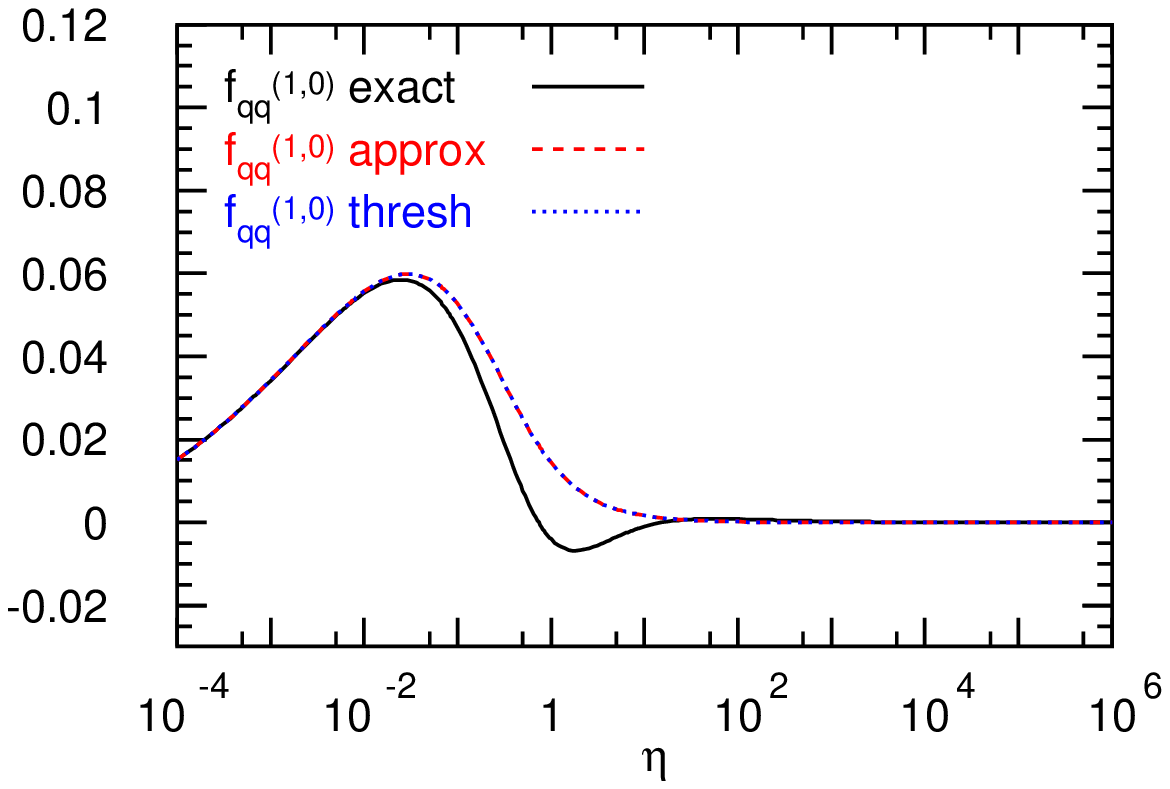}
    \includegraphics[width=7.25cm]{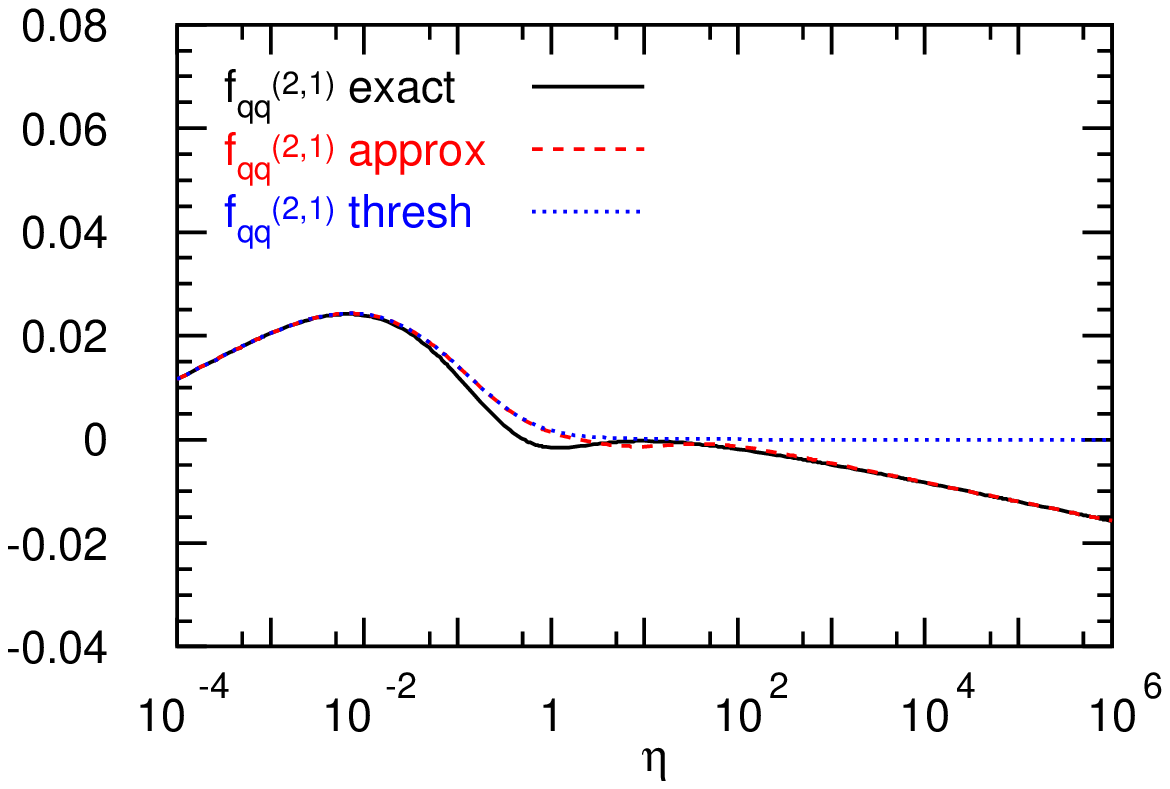}
    \includegraphics[width=7.25cm]{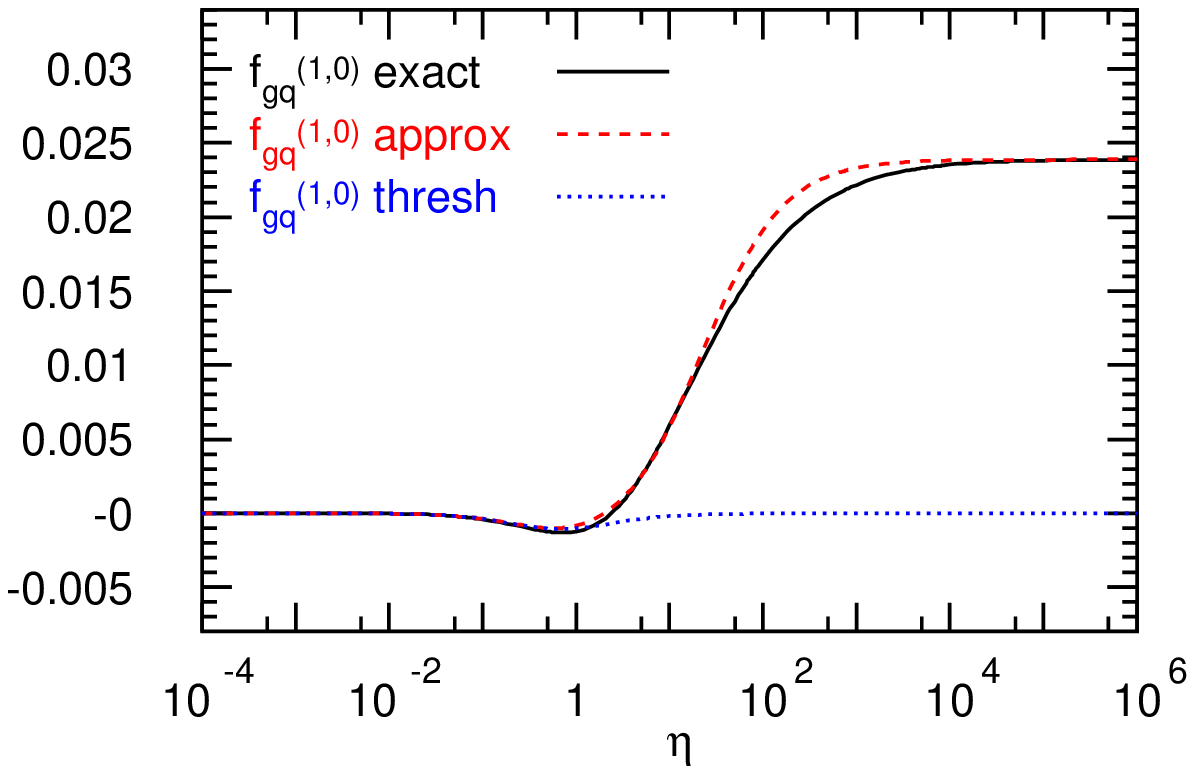}
    \includegraphics[width=7.25cm]{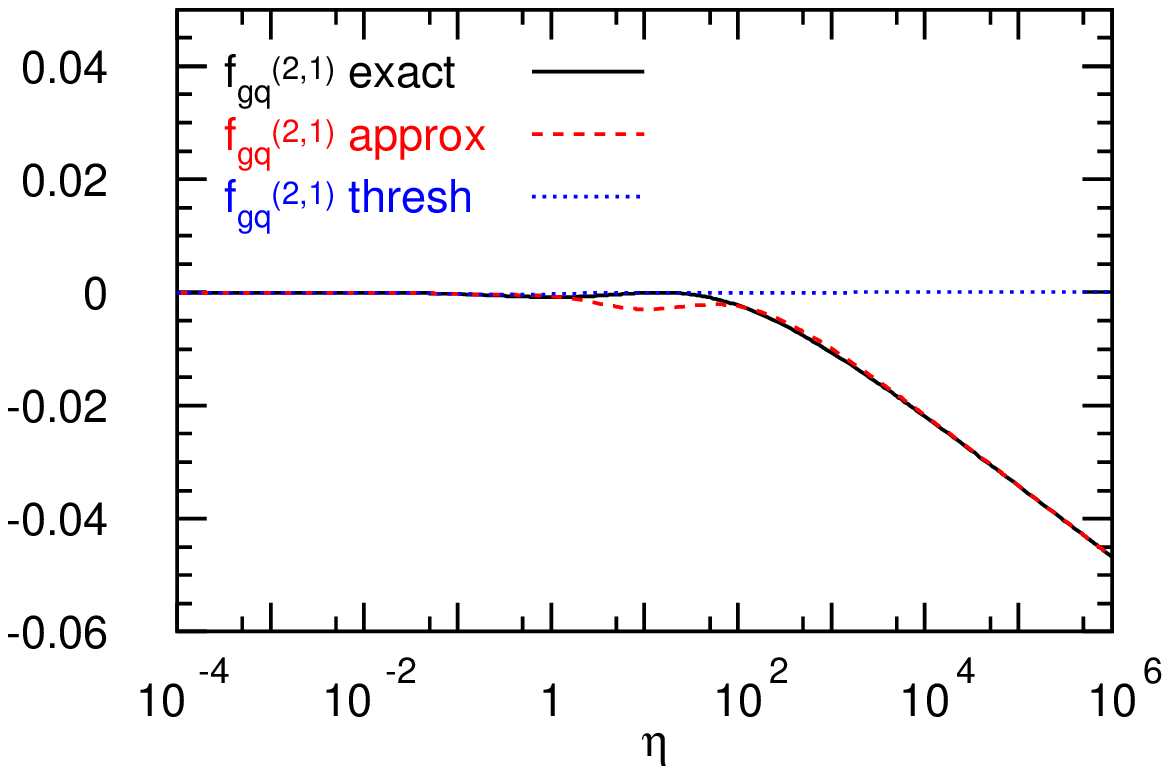}
    \includegraphics[width=7.25cm]{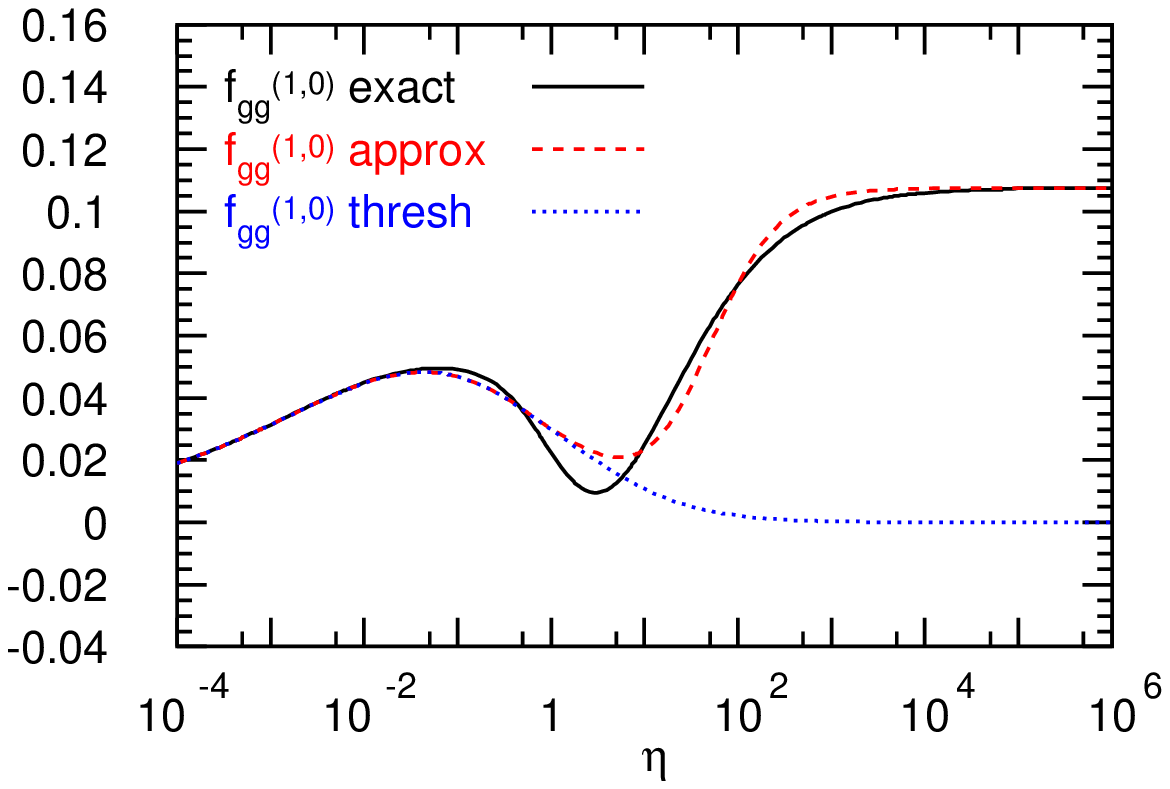}
    \includegraphics[width=7.25cm]{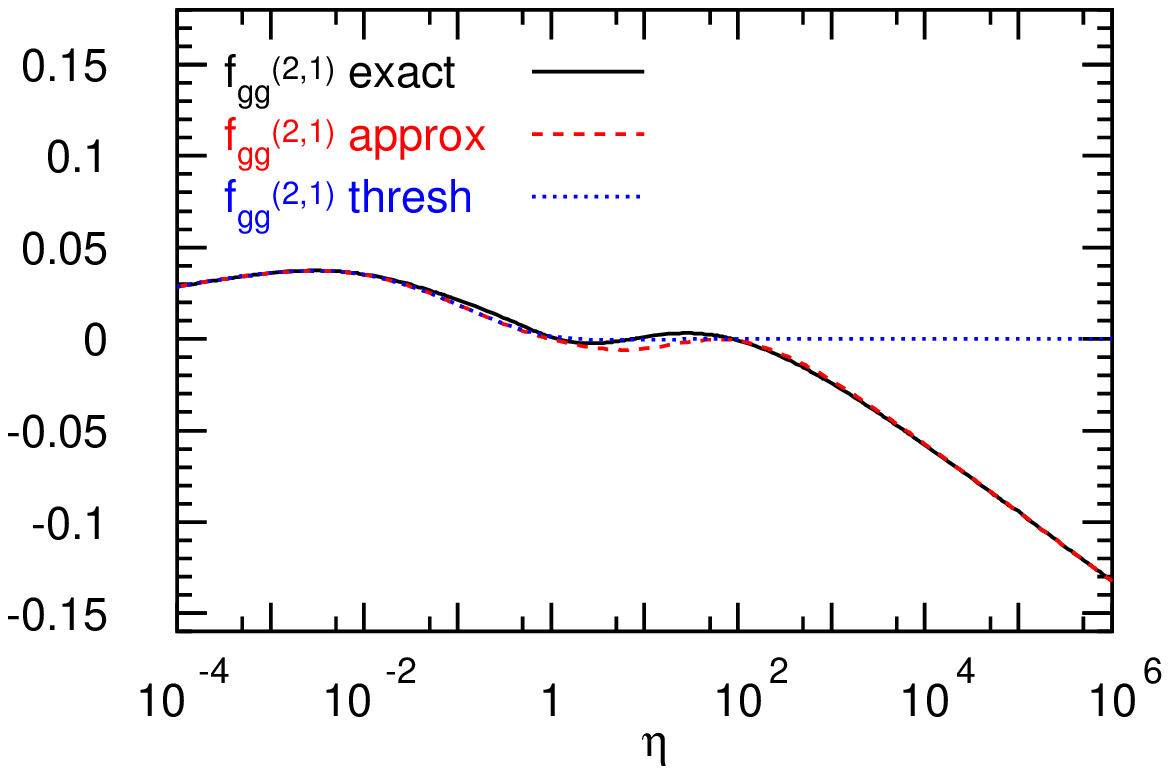}
    }
\vspace*{-1mm}
    \caption{ \small
      \label{fig:fij}
      Comparison of exact results for $\fijon$ and $\fijto$ with the threshold expansions and the
      approximations defined in eqs.~(\ref{eq:assembly1}) and
      (\ref{eq:assembly2}).
    }
\end{figure}
Let us now employ the above findings. 
Specifically, we are interested in combining the approximations in the
two kinematical regions, i.e., at threshold and at high energy (small-$x$) 
in order to arrive at smoothly interpolating functional forms for the scaling functions.
Whenever possible, we compare to the exact results in order to check the
quality of the approach.
We choose the following ansatz for $f_{ij}^{(l)}$ at one- and two-loops,
\begin{eqnarray}
  \label{eq:assembly1}
  f_{ij}^{(1)} &=& 
  \rho^l\, f_{ij}^{(1){\rm thresh}}
  + \beta^k\, f_{ij}^{(1){\rm LL}x} \,\frac{\eta^{\gamma}}{C + \eta^{\gamma}} 
  \, ,
\\
  \label{eq:assembly2}
  f_{ij}^{(2)} &=& 
  \rho^l\, f_{ij}^{(2){\rm thresh}}
  + \beta^k\, \biggl( 
  - \ln \rho\, f_{ij}^{(2){\rm LL}x} 
  + f_{ij}^{(2){\rm NLL}x} \,\frac{\eta^{\gamma}}{C + \eta^{\gamma}} 
  \biggr)
  \, ,
\end{eqnarray}
where $\beta = \sqrt{1-\rho}$ is the heavy-quark velocity and 
$\eta = (1/\rho-1)$ is the distance from threshold.
For the parton channels $ij=q{\bar q},gg$ the parameters $k,l$ 
take the values $k=3$, $l=0$ and for $ij=gq$ we have $k=5$, $l=1$.
These values reflect the exact functional dependence on $\beta$ and $\rho$ 
in the respective kinematical limits.
The well-known threshold expansions are denoted $f_{ij}^{(l){\rm thresh}}$ and given, e.g., in~\cite{Aliev:2010zk}. 
The high-energy asymptotic behavior is split in LL$_x$ and NLL$_x$ parts 
$f_{ij}^{(l){\rm LL}x}$ and $f_{ij}^{(l){\rm NLL}x}$ 
corresponding to eqs.~(\ref{eq:fqq10})--(\ref{eq:f22gg}).
The high-$\eta$ tail proportional to ${\rm const}_\rho$ 
(or $1/\omega$ in Mellin space) 
is smoothly matched with a factor $\eta^{\gamma}/(C + \eta^{\gamma})$. 
The suppression parameters $\gamma, C$ in eqs.~(\ref{eq:assembly1}), (\ref{eq:assembly2}) 
take the following values at NLO as a best fit for $\fijon$,
\begin{equation}
  \label{eq:suppr-param10}
  \gamma = 0.99\, ,~~ C = 20.9
  ~~ {\rm for}~ gq \qquad {\rm and}\qquad
  \gamma = 1.18\, ,~~ C = 97.3
  ~~ {\rm for}~gg
  \, ,
\end{equation}
and at NNLO fitted to $\fijto$,
\begin{equation}
  \label{eq:suppr-param21}
  \gamma = 1.37\, ,~~ C = 47.9
  ~~ {\rm for}~ q{\bar q} \, , \quad
  \gamma = 0.90\, ,~~ C = 16.4
  ~~ {\rm for}~ gq \quad {\rm and}\quad
  \gamma = 0.84\, ,~~ C = 12.6
  ~~ {\rm for}~gg
  \, .
\end{equation}

In Fig.~\ref{fig:fij} we show the results of this procedure for the scaling functions $\fijon$ and $\fijto$. 
In particular, we compare the exact results with the approximations of 
eqs.~(\ref{eq:assembly1}), (\ref{eq:assembly2}) using the values of
eqs.~(\ref{eq:suppr-param10}) and (\ref{eq:suppr-param21}) for the 
parameters $\gamma$ and $C$.
The plots in Fig.~\ref{fig:fij} 
show a perfect match at both end of the kinematical range with an accuracy at the per mille
level and even better as $s \to 4m^2$ and for $s \gg m^2$. 
This is very a strong check in particular on the results of $f_{ij}^{(21)}$ which 
are known numerically from renormalization group methods~\cite{Langenfeld:2009wd}. 
Some deviations between the approximations of eqs.~(\ref{eq:assembly1}), (\ref{eq:assembly2})
and the exact results in the central range of $\eta \approx 0.1 \dots 10$ are visible in Fig.~\ref{fig:fij}.
However, these have generally a small impact on cross section predictions for
hadron colliders, because the necessary convolution with the parton luminosities in 
eq.~(\ref{eq:totalcrs}) is a non-local operation and has a smoothening effect.
Moreover, the parton luminosities are steeply falling functions as $\eta$ grows large, 
giving numerically the most weight to the threshold region, which is after all 
the rational behind phenomenology based on the threshold resummation.
In summary, the plots in Fig.~\ref{fig:fij} demonstrate that the chosen
approach of combining the threshold expansion and the high-energy
asymptotics leads to very good approximations of the exact scaling functions.

\begin{figure}[th!]
\centering
    {
    \includegraphics[width=7.25cm]{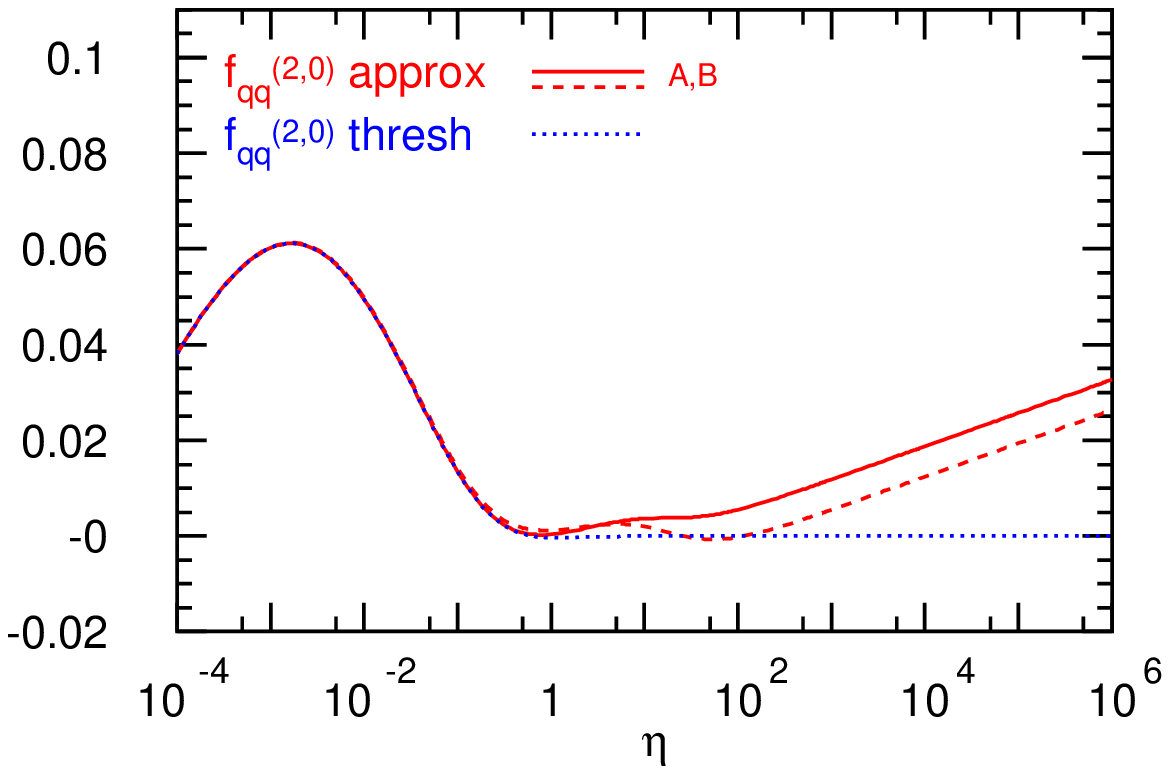}
    \includegraphics[width=7.25cm]{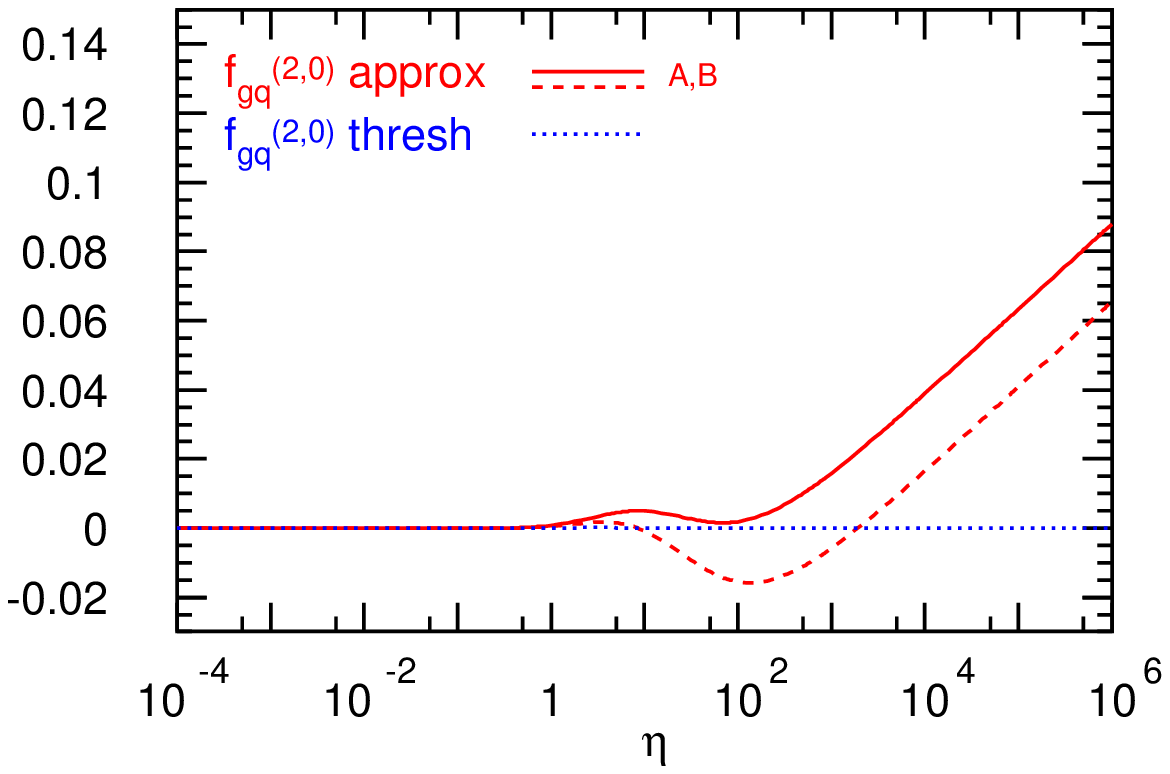}
    \includegraphics[width=7.25cm]{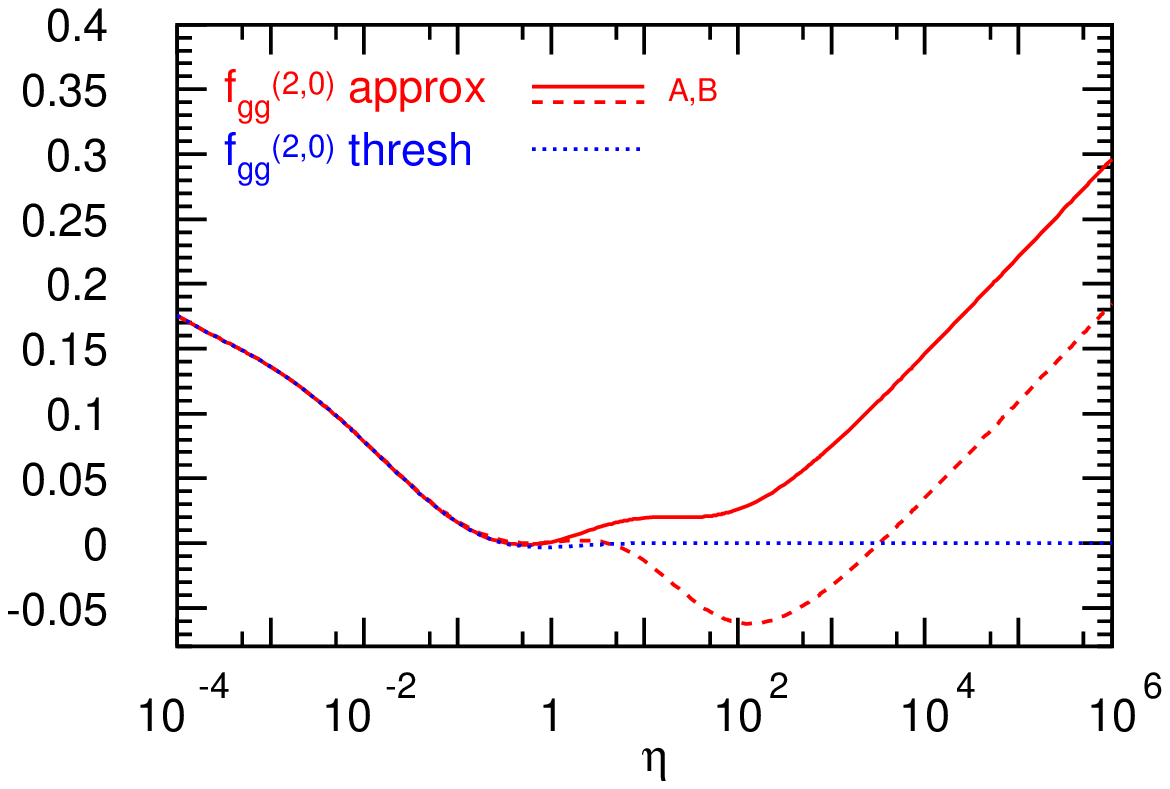}
    }
\vspace*{-1mm}
    \caption{ \small
      \label{fig:f20}
The threshold expansions for $\fijtn$ and the approximations defined in 
eqs.~(\ref{eq:assemblyij20}) and (\ref{eq:assemblygq20}). 
The two curves (solid=A, dashed=B) correspond to the choices for the constants
given in eqs.~(\ref{eq:qqAB})--(\ref{eq:ggAB}).
    }
\end{figure}
The main object of our interest are the scaling functions $\fijtn$. 
Here we aim at defining an uncertainty band which combines both, 
the threshold approximation and the high-energy limit, 
and also accounts for an error estimate due to the uncalculated 
next term in the expansions in either kinematical region. 
At large $\eta$, this is achieved with the NLL$_x$ terms in $\fqqtn$ $\fgqtn$ and $\fggtn$ 
which contain the values of $\cqqtn$, $\cgqtn$ and $\cggtn$ 
with the conservatively estimated ranges given in eqs.~(\ref{eq:cgq20})--(\ref{eq:cqq20}). 
The known threshold contributions for the functions $f_{qq}^{(20){\rm thresh}}$ and $f_{gg}^{(20){\rm thresh}}$ 
on the other hand contain the complete tower of logarithmically enhanced terms 
in $\ln^k \beta$, where $k=1,\dots,4$, 
as well as all Coulomb corrections at two loops proportional to 
$1/\beta^2$ and $1/\beta$ which dominate as $\beta \to 0$.
Therefore, an estimate for an additional contribution of order 
${\cal O}({\rm const}_\beta)$ (and vanishing as $\rho \to 1$) 
to be added to $f_{qq}^{(20){\rm thresh}}$ and $f_{gg}^{(20){\rm thresh}}$ 
serves as check on their inherent uncertainty. 
A $[0,1]$ Pad{\'e} estimate based on the exact NLO results $\fqqon$ and $\fggon$ 
yields for these constant $\cbqqtn$ and $\cbggtn$ 
in the normalization of eq.~(\ref{eq:DefinitionScalingFunctions}) 
the values,
\begin{eqnarray}
  \label{eq:constqq-beta}
\cbqqtn &=& \frac{\fqqn}{(4\pi)^4} 
\left( 
     {1276 \over 9} - 172\*\lntwo + {256\over 3}\*\lnsqtwo 
     - {86 \over 3}\*\ztwo - {20 \over 9}\*\nf + {8 \over 3}\*\nf\*\lntwo
\right)^2\, ,
\\
  \label{eq:constgg-beta}
\cbggtn &=& \frac{\fqqn}{(4\pi)^4} 
\left( 
     {4444 \over 21} - {2136 \over 7}\*\lntwo + 192\*\lnsqtwo - {283 \over 7}\*\ztwo 
\right)^2\, ,
\end{eqnarray}
while the default values in phenomenological studies 
are usually taken as $\cbqqtn = \cbggtn = 0$, 
see, e.g., the discussion in \cite{Aliev:2010zk}.
We neglect the $gq$-channel in these considerations, since it is 
very small near threshold anyway.

Thus, on the basis of eq.~(\ref{eq:assembly2}) 
and the discussion above we take the following two variants for the unknown
full $\rho$ and $\eta$ dependence of the two-loop scaling functions,
\begin{eqnarray}
  \label{eq:assemblyij20}
  f_{ij}^{(20)A/B} &=& 
  f_{ij}^{(20){\rm thresh}} + C_{\beta,ij}^{(20)A/B} 
  + \beta^3\, f_{ij}^{(2){\rm LL}x}\, \biggl( - \ln \rho\,   
  + r_{x,ij}^{A/B} \,\frac{\eta^{\gamma}}{C + \eta^{\gamma}} 
  \biggr)
  \, ,
  \quad
      {\rm for}~ ij=qq, gg
  \, ,
  \quad
\\
  \label{eq:assemblygq20}
  f_{gq}^{(20)A/B} &=& 
  \rho\, 
  f_{gq}^{(20){\rm thresh}}
  + \beta^5\, f_{gq}^{(2){\rm LL}x}\, \biggl( - \ln \rho\,   
  + r_{x,gq}^{A/B} \,\frac{\eta^{\gamma}}{C + \eta^{\gamma}} 
  \biggr)
  \, ,
\end{eqnarray}
where we take the same parameters $\gamma$ and $C$ for the respective channel as determined for 
$\fijto$ in eq.~(\ref{eq:suppr-param21}) and the values for 
$r_{x,ij}$ and $C_{\beta,ij}^{(20)}$ are chosen as 
\begin{eqnarray}
  \label{eq:qqAB}
  C_{\beta,qq}^{(20)A} \,=\, 0
  \, ,
  \quad  
  r_{x,qq}^{A} \,=\, - 3.0
  \, ,
  &\qquad\qquad&  
  C_{\beta,qq}^{(20)B} \,=\, \cbqqtn
  \, ,
  \quad  
  r_{x,qq}^{B} \,=\, -5.1
  \, ,
\\
  \label{eq:gqAB}
  r_{x,gq}^{A} \,=\, - 5.6
  \, ,
  &\qquad\qquad&  
  \phantom{C_{\beta,qq}^{(20)B} \,=\, \cbqqtn
  \, ,
  \quad}
  r_{x,gq}^{B} \,=\, -7.7
  \, ,
\\
  \label{eq:ggAB}
  C_{\beta,gg}^{(20)A} \,=\, 0
  \, ,
  \quad  
  r_{x,gg}^{A} \,=\, - 4.8
  \, ,
  &\qquad\qquad&  
  C_{\beta,gg}^{(20)B} \,=\, \cbggtn
  \, ,
  \quad  
  r_{x,gg}^{B} \,=\, -8.2
  \, .
\end{eqnarray}

The results for eqs.~(\ref{eq:assemblyij20}) and (\ref{eq:assemblygq20}) are displayed in Fig.~\ref{fig:f20}. 
The above procedure leads to the bands shown which widen significantly for large center-of-mass energies 
and rise with the same slope as $s \gg m^2$ 
due to the known logarithmic dependence on $\rho$ of the LL$_x$ terms.
It is evident from Fig.~\ref{fig:f20} and the numerical size of the 
various constants, $\cbqqtn$ and $\cbggtn$ in eqs.~(\ref{eq:constqq-beta}), (\ref{eq:constgg-beta}) 
as well as $\cqqtn$, $\cgqtn$ and $\cggtn$ in eqs.~(\ref{eq:cgq20})--(\ref{eq:cqq20}) 
that the uncertainty in the latter is dominating even in the range of $\eta = 1 \dots 100$. 
Therefore a more accurate determination of $\cqqtn$, $\cgqtn$ and $\cggtn$,  
preferably a computation from first principles, is highly desirable.
To a minor extent, the bands in Fig.~\ref{fig:f20} depend on the chosen matching, i.e.,
on eq.~(\ref{eq:suppr-param21}). 
However, the values for $\gamma$ and $C$ in eq.~(\ref{eq:suppr-param21}) are all of the same order 
and, as we have shown in Fig.~\ref{fig:fij} this part of our procedure 
leads to reasonable descriptions in all cases where exact results are available.

\bigskip

\begin{table}[th!]
\renewcommand{\arraystretch}{1.3}
\begin{center}
{\small
\hspace*{-5mm}
\begin{tabular}{|l|l|l|l|l|l|}
\hline
&\multicolumn{1}{|c|}{TEV $\sqrt{S} = 1.96\, {\rm TeV}$ } 
&\multicolumn{1}{|c|}{LHC $\sqrt{S} = 7\, {\rm TeV}$ }
&\multicolumn{1}{|c|}{LHC $\sqrt{S} = 8\, {\rm TeV}$ }
&\multicolumn{1}{|c|}{LHC $\sqrt{S} = 14\, {\rm TeV}$ }
\\     
\hline
thresh &
$6.90~^{+0.26}_{-0.32}~^{+0.16}_{-0.16}$&
$130.4~^{+2.9}_{-7.2}~^{+5.9}_{-5.9}$&
$190.5~^{+3.7}_{-10.2}~^{+8.0}_{-8.0}$&
$795.3~^{+9.0}_{-35.0}~^{+23.3}_{-23.3}$
\\
(A+B)/2& 
$7.01~^{+0.34}_{-0.37}~^{+0.16}_{-0.16}~(^{+0.03}_{-0.03})$&
$138.5~^{+8.1}_{-10.2}~^{+6.4}_{-6.4}~(^{+3.1}_{-3.1})$&
$202.5~^{+11.3}_{-14.5}~^{+8.6}_{-8.6}~(^{+5.2}_{-5.2})$&
$845.5~^{+37.3}_{-51.9}~^{+25.3}_{-25.3}~(^{+34.2}_{-34.2})$
\\
\hline
\end{tabular}
}
\caption{\small 
The total cross section for top-quark pair-production at (approximate) NNLO using a 
pole mass $\mt = 173~{\rm GeV}$ and the ABM11 PDF set~\cite{Alekhin:2012ig} 
with errors shown as 
$\sigma + \Delta \sigma_{\rm scale} + \Delta \sigma_{\rm PDF} (+ \Delta \sigma_{\rm A/B})$
The scale uncertainty $ \Delta \sigma_{\rm scale}$ is based on maximal and minimal 
shifts for the choices $\mu=\mt/2$ and $\mu = 2\mt$,  
$\Delta \sigma_{\rm PDF}$ is the 1$\sigma$ combined PDF+$\alpha_s$ error 
and the $\Delta \sigma_{\rm A/B}$ is the deviation of the central value for either variant A or B
of eqs.~(\ref{eq:assemblyij20}) and (\ref{eq:assemblygq20}).
All rates are in pb. 
}
\label{tab:ttbar-abm11}
\end{center}
\end{table}

\begin{table}[th!]
\renewcommand{\arraystretch}{1.3}
\begin{center}
{\small
\hspace*{-5mm}
\begin{tabular}{|l|l|l|l|l|l|}
\hline
&\multicolumn{1}{|c|}{TEV $\sqrt{S} = 1.96\, {\rm TeV}$ } 
&\multicolumn{1}{|c|}{LHC $\sqrt{S} = 7\, {\rm TeV}$ }
&\multicolumn{1}{|c|}{LHC $\sqrt{S} = 8\, {\rm TeV}$ }
&\multicolumn{1}{|c|}{LHC $\sqrt{S} = 14\, {\rm TeV}$ }
\\     
\hline
thresh &
$7.10~^{+0.16}_{-0.14}~^{+0.15}_{-0.15}$&
$135.4~^{+0.0}_{-3.1}~^{+5.9}_{-5.9}$&
$197.6~^{+0.0}_{-4.9}~^{+7.9}_{-7.9}$&
$820.5~^{+0.0}_{-24.0} ~^{+22.4}_{-22.4}$
\\
(A+B)/2& 
$7.26~^{+0.28}_{-0.20}~^{+0.15}_{-0.15}~(^{+0.04}_{-0.04})$&
$146.4~^{+4.0}_{-6.9}~^{+6.5}_{-6.5}~(^{+4.4}_{-4.4})$&
$213.7~^{+5.3}_{-9.7}~^{+8.7}_{-8.7}~(^{+7.3}_{-7.3})$&
$885.7~^{+12.3}_{-33.4} ~^{+25.0}_{-25.0}~(^{+46.3}_{-46.3})$
\\
\hline
\end{tabular}
}
\caption{\small 
Same as Tab.~\ref{tab:ttbar-abm11} for a running mass 
$\mt(\mt) = 164~{\rm GeV}$ in the $\overline{MS}$ scheme.
}
\label{tab:ttbar-abm11-msbar}
\end{center}
\end{table}

\begin{table}[th!]
\renewcommand{\arraystretch}{1.3}
\begin{center}
{\small
\hspace*{-5mm}
\begin{tabular}{|l|l|l|l|l|l|}
\hline
&\multicolumn{1}{|c|}{TEV $\sqrt{S} = 1.96\, {\rm TeV}$ } 
&\multicolumn{1}{|c|}{LHC $\sqrt{S} = 7\, {\rm TeV}$ }
&\multicolumn{1}{|c|}{LHC $\sqrt{S} = 8\, {\rm TeV}$ }
&\multicolumn{1}{|c|}{LHC $\sqrt{S} = 14\, {\rm TeV}$ }
\\     
\hline
thresh &
$7.13~^{+0.30}_{-0.40}~^{+0.17}_{-0.12}$&
$164.3~^{+3.3}_{-9.2}~^{+4.4}_{-4.5}$&
$234.6~^{+4.1}_{-12.6}~^{+5.8}_{-5.9}$&
$908.2~^{+9.9}_{-40.6}~^{+15.2}_{-16.7}$
\\
(A+B)/2 & 
$7.27~^{+0.41}_{-0.46}~^{+0.17}_{-0.12}~(^{+0.03}_{-0.03})$&
$174.9~^{+10.3}_{-13.2}~^{+4.7}_{-4.8}~(^{+4.6}_{-4.6})$&
$249.9~^{+14.0}_{-18.2}~^{+6.2}_{-6.3}~(^{+7.5}_{-7.5})$&
$967.2~^{+43.0}_{-60.3}~^{+16.0}_{-17.6}~(^{+44.9}_{-44.9})$
\\
\hline
\end{tabular}
}
\caption{\small 
Same as Tab.~\ref{tab:ttbar-abm11} for the MSTW PDF set~\cite{Martin:2009iq}.  
}
\label{tab:ttbar-mstw}
\end{center}
\end{table}

Let us finally investigate the implications 
for the total cross sections of top-quark pair-production at Tevatron and the LHC.
To that end, we have implemented the approximate scaling functions $\fqqtn$, $\fgqtn$ and $\fggtn$  
as defined in eqs.~(\ref{eq:assemblyij20}) and (\ref{eq:assemblygq20}) 
in a new version of the program {\sc Hathor}~\cite{Aliev:2010zk}, 
which otherwise uses the exact results for the scaling functions 
at NLO as well as for $\fijto$ and $\fijtt$.
Any difference that would arise from using the approximate expression 
eq.~(\ref{eq:assembly2}) for the scaling functions $\fijto$ 
instead is marginal, cf. Fig.~\ref{fig:fij}.

As a central prediction we take the average of the two variants defined 
in eqs.~(\ref{eq:assemblyij20}) and (\ref{eq:assemblygq20}) 
which we denote as '(A+B)/2' and compare to previous estimates 
of~\cite{Langenfeld:2009wd,Aliev:2010zk} based on threshold approximation and labeled as 'thresh'.
The new NNLO approximation accounts for the effect of all parton channels which are
also non-zero at NLO, including the $gq$-channel, which picks up some
contribution of the high-energy region at NNLO.
However, we neglect any effect of new parton channels at NNLO, 
i.e., $qq$, ${\bar q}\,{\bar q}$ and $q_i{\bar q}_j$ (for unlike flavors $i \neq j$). 
The theoretical uncertainty is determined from the scale variation considering
the choices $\mu = \mt/2$ and $\mu = 2\mt$ 
and taking the maximum and minimum of respective shifts of the cross sections. 
This is sufficient since at NNLO the $\muf$ dependence is generally weak and
the scale uncertainty is mainly driven by the $\mur$ variation.

Choosing a pole mass of $\mt = 173~{\rm GeV}$ and the ABM11 PDF set~\cite{Alekhin:2012ig} 
at NNLO our predictions are shown in Tab.~\ref{tab:ttbar-abm11}.
Comparing the threshold approximation and the new estimate '(A+B)/2' we see
that there are generally small positive shifts in the cross sections due to the high-energy tail. 
As expected, the effect of the high-energy limit is rather modest, 
which nicely illustrates and confirms the stability of predictions based on
soft gluon resummation. 
We note that the small shifts in the central values of the cross section predictions
are in line with the inherent uncertainty attributed to previous approximations~\cite{Langenfeld:2009wd}.
For the Tevatron these amount to ${\cal O}(1-2\%)$ consistent with the previously observed small effect 
of hard radiation (not accounted for by threshold resummation) 
on the total cross section of $t{\bar t}$+jet production~\cite{Dittmaier:2007wz,Dittmaier:2008uj,Melnikov:2009dn}.
For the LHC at all center-of-mass energies $\sqrt{S} = 7, 8$ and  $ 14\, {\rm TeV}$
these shifts are larger of the order ${\cal O}(6-8\%)$ 
due to the parton luminosities $\Lij$ giving more weight to the 
(positive) high-energy tail of all scaling functions in eq.~(\ref{eq:totalcrs}).
The scale uncertainties in the new estimate '(A+B)/2' increase compared to previous 
analyses -- sometimes by up to a factor of two. 
To a large extent this increase is due to the $gq$-channel, where the high-energy tail 
is numerically more important than the threshold region, cf. Fig.~\ref{fig:f20}.
Taking, e.g., the values at $\sqrt{S} = 7\, {\rm TeV}$ in Tab.~\ref{tab:ttbar-abm11}, 
the cross sections split up into the contributions of the individual channels as
\begin{eqnarray}
\sigma_{(A+B)/2}(\mu=m/2) 
&=\, 146.6 {\rm pb} 
&=\, (97.4 {\rm pb})_{\rm gg} 
+ (27.9 {\rm pb})_{\rm q \bar q} 
+ (21.3 {\rm pb})_{\rm qg}\, ,
\\
\sigma_{(A+B)/2}(\mu=m) 
&=\, 138.5 {\rm pb} 
&=\, (106.4 {\rm pb})_{\rm gg} 
+ (28.3 {\rm pb})_{\rm q \bar q} 
+ (3.8 {\rm pb})_{\rm qg}\, ,
\\
\sigma_{(A+B)/2}(\mu=2m) 
&=\, 128.3 {\rm pb} 
&=\, (108.5 {\rm pb})_{\rm gg} 
+ (28.5 {\rm pb})_{\rm q \bar q} 
+ (-8.7 {\rm pb})_{\rm qg} \, .
\end{eqnarray}
The relatively larger impact of the $qg$ channel can be understood from the fact 
that the scale dependence of its high-energy tail is not entirely compensated 
at the accuracy given here and, 
thus, leads to an increase of the scale uncertainty compared to earlier studies. 
This also underlines the importance of considering the high-energy regime, ignored in previous
analyses~\cite{Langenfeld:2009wd,Ahrens:2010zv,Kidonakis:2010dk,Beneke:2011mq,Cacciari:2011hy}, 
for all LHC predictions.
Compared to NLO predictions however, we still observe a significant improvement.
For the new estimate '(A+B)/2' we also quote the systematic uncertainty 
from choosing either variant A or B in eq.~(\ref{eq:assemblyij20}) and (\ref{eq:assemblygq20}).
We see for all cases that those shifts are comparable to or smaller than the
scale uncertainty.

In Tab.~\ref{tab:ttbar-abm11-msbar} we repeat the computation for the corresponding running mass
of $\mt = 164~{\rm GeV}$ in the $\overline{MS}$ scheme and similar conclusions
hold with respect to the pattern of observed changes.
In particular, we note that in this mass scheme better scale stability is achieved, 
corroborating the findings of \cite{Langenfeld:2009wd}.
This implies that the NNLO approximation uncertainty is considerably larger
than the scale dependence for the case of LHC with $\sqrt{S} = 14\, {\rm TeV}$.
Finally in Tab.~\ref{tab:ttbar-mstw} we choose the MSTW PDF set~\cite{Martin:2009iq} for comparison.
While the Tevatron predictions of both sets are largely in agreement, 
the difference in the LHC predictions can be attributed to differences in the
parametrization of the gluon PDFs at moderately large $x$ and different central values for $\alpha_s$, 
see also~\cite{Alekhin:2012ig} for more PDF comparisons.

In summary we present a phenomenological study of heavy-quark hadro-production 
including new results in the high-energy limit as $s \gg m^2$. 
We have provided approximate NNLO QCD corrections for the full kinematic range
based on those new constraints from high-energy factorization 
combined with existing results for the threshold region for $s \gsim 4m^2$.
Our investigations have quantified the largest residual uncertainty 
in the two-loop scaling functions at large $\eta$ due to our incomplete 
knowledge of the subleading `small-$x$' terms in $\fijtn$.
In view of this it is therefore an important task to compute 
the exact result for those NLL$_x$ terms, $\fijtn$ in eq.~(\ref{eq:assembly2}), 
e.g., following~\cite{Catani:1990eg,Ball:2001pq} 
or by using the available NNLO QCD predictions for heavy-quark hadro-production 
in the small mass limit~\cite{Czakon:2007ej,Czakon:2007wk}.
This would immediately remove the major source of the current residual uncertainty.
Other improvements of the theoretical accuracy relying on generalizations of  
the resummations at threshold and high-energy, 
e.g., along the lines of~\cite{Moch:2009hr,Soar:2009yh} and~\cite{Vogt:2011jv}, 
can be considered as well.

For the predictions of the total cross section of top-quark pair-production 
at Tevatron and the LHC the current available information leads to uncertainties 
in the approximate NNLO results which are of the order 
4\% for Tevatron and 5 \% for the LHC.
Further important corrections to be considered in phenomenological studies 
and not accounted for here arise from the electro-weak radiative corrections 
at NLO~\cite{Beenakker:1993yr,Bernreuther:2006vg,Kuhn:2006vh} 
as well as from bound state effects and the resummation 
of Coulomb type corrections~\cite{Hagiwara:2008df,Kiyo:2008bv,Beneke:2011mq}. 

At present, our approximate results represent the most complete NNLO predictions 
for heavy-quark hadro-production. 
The phenomenological importance of this process motivates further improvements to 
reduce the theoretical uncertainty and a number of directions for future
research have been proposed.
The improved QCD corrections have been implemented 
in a new version of the program {\sc Hathor}~\cite{Aliev:2010zk} 
which is publicly available for download from~\cite{hathor:2010} or from the authors upon request.

\subsection*{Note added:} 
While we were finishing this paper, numerically determined complete results 
for the parton channel $q\bar q \to t \bar t$ at NNLO have been presented 
in~\cite{Baernreuther:2012ws} including 
the double real emission $q\bar q \to t \bar t q^\prime\bar q^\prime$ for $q \neq q^\prime$.
These numerical results are not sufficient for a comparison in the high-energy region $\eta \ge 100$, 
where the parton channel with double real emission $q \bar q \to t \bar t q \bar q$ dominates.
However, the results are consistent for smaller $\eta$ values.
Also the NNLO result of \cite{Baernreuther:2012ws} is included 
as an option in the new version of {\sc Hathor}~\cite{Aliev:2010zk}. 

\subsection*{Acknowledgments}
We thank R.K. Ellis and A. Sabio Vera for discussions. 
We have used the latest version of {\tt FORM}~\cite{Vermaseren:2000nd} for the analytic calculations.
This work is partially supported 
by the Deutsche Forschungsgemeinschaft in Sonderforschungs\-be\-reich/Transregio~9,
by Helmholtz Gemeinschaft under contract VH-HA-101 ({\it Alliance Physics at the Tera\-scale}), 
by the UK Science \& Technology Facilities Council under grant number ST/G00062X/1 
and by the European Commission through contract PITN-GA-2010-264564 ({\it LHCPhenoNet}).

%
%
{\footnotesize


}

\end{document}